\documentclass[]{aa}
\usepackage{graphicx}
\usepackage{amsmath}
\usepackage{amssymb}
\usepackage{natbib}
\usepackage{color}

\begin{document}

\newcommand{\czero}{\element{C}}
\newcommand{\ci}{[\ion{C}{i}]}
\newcommand{\cii}{[\ion{C}{ii}]}
\newcommand{\co}{\element{CO}}
\newcommand{\col}[1]{\ensuremath{N_\mathrm{#1}}}
\newcommand{\cplus}{\element[+]{C}}
\newcommand{\cmark}[1]{\textcolor{red}{#1}}
\newcommand{\cms}{clump-mass spectrum}
\newcommand{\diff}[1]{\ensuremath{\mathrm{d}#1}}
\newcommand{\ddiff}[2]{\frac{\ensuremath{\mathrm{d}#1}}{\ensuremath{\mathrm{d}#2}}}
\newcommand{\ergpspsc}{\ensuremath{\mathrm{erg}\ \mathrm{s}^{-1}\ \mathrm{cm}^{-2}}}
\newcommand{\fdraine}{\ensuremath{\chi_\mathrm{D}}}
\newcommand{\hi}{\ion{H}{i}}
\newcommand{\hii}{\ion{H}{ii}}
\newcommand{\htwo}{\ensuremath{\element{H}_2}}
\newcommand{\ind}[1]{\ensuremath{_\mathrm{#1}}}
\newcommand{\ism}{interstellar medium}
\newcommand{\kosma}{KOSMA-$\tau$ PDR-model}
\newcommand{\msr}{mass-size relation}
\newcommand{\msun}{\ensuremath{\mathrm{M}_\mathrm{\sun}}}
\newcommand{\mum}{\ensuremath{\mu\mathrm{m}}}
\newcommand{\nii}{[\ion{N}{ii}]}
\newcommand{\ozero}{\element{O}}
\newcommand{\oi}{[\ion{O}{i}]}
\newcommand{\oiii}{[\ion{O}{iii}]}
\newcommand{\pcc}{\ensuremath{\mathrm{cm}^{-3}}}
\newcommand{\ps}{\ensuremath{\mathrm{s}^{-1}}}
\newcommand{\psc}{\ensuremath{\mathrm{cm}^{-2}}}
\newcommand{\thirteenc}{\element[][13]{C}}
\newcommand{\thirteenco}{\element[][13]{CO}}
\newcommand{\twelveco}{\element[][12]{CO}}

   \title{
A clumpy-cloud PDR model of the global far-infrared line emission of the Milky Way
}

   \author{M. Cubick \inst{1}
          \and
          J. Stutzki\inst{1}
          \and
          V. Ossenkopf \inst{1,2}
          \and
          C. Kramer \inst{1} 
          \and
          M. R\"ollig \inst{1,3}
          }

   \offprints{M. Cubick}

   \institute{I. Physikalisches Institut, Universit\"at zu K\"oln,
              Z\"ulpicher Str. 77, D-50937 K\"oln, Germany\\
              \email{cubick@ph1.uni-koeln.de}
             \and
              SRON Netherlands Institute for Space Research, 
              PO Box 800, 9700 AV Groningen, The Netherlands\\
             \and
              Argelander Institut f\"ur Radioastronomie, Universit\"at Bonn,
              Auf dem H\"ugel 71, D-53121 Bonn, Germany\\
             }

   \date{Received \today; accepted }

 
  \abstract
   {
The fractal structure of the interstellar medium suggests that
the interaction of UV radiation with the ISM as described in
the context of photon-dominated regions (PDR) dominates most of the
physical and chemical conditions, and hence the far-infrared and submm
emission from the ISM in the Milky Way.
}
   {
We investigate to what extent the Galactic FIR line emission of
the important species \co, \czero, \cplus, and \ozero, 
as observed by the Cosmic Background Explorer (COBE) satellite can be 
modeled in the framework of a clumpy, UV-penetrated cloud scenario.
}
   {
The far-infrared line emission of the Milky Way 
is modeled as the emission 
from an ensemble of clumps with a power law 
clump mass spectrum and mass-size
relation with power-law indices consistent with the observed ISM structure. 
The individual clump line intensities are calculated using
the \kosma{} for spherical clumps. 
The model parameters for the cylindrically symmetric Galactic distribution of 
the mass density and volume filling factor are determined by the observed radial
distributions. A constant FUV intensity, in which the clumps are embedded, is assumed.
}
   {
We show that this scenario can explain, without any further
assumptions and within a factor of about 2,
the absolute FIR-line intensities and their distribution with
Galactic longitude as observed by COBE.
}
   {}


   \maketitle
%

\section{Introduction}
\label{s_intro}

The importance of the interaction between the interstellar
UV-radiation and dense clouds in the ISM, determining the 
physical and chemical conditions in the 
surface regions of molecular clouds, which  
are modelled as so-called photon-dominated
regions (PDRs), has been recognized since the first observations of
the dominant cooling lines in the FIR, \cii~158 \mum, 
\citep{Russell1980, Russell1981, Stutzki1988, Mizutani1994}, 
\oi~63 \mum{} \citep{Melnick1979, Stacey1983}
and the mid- and high-$J$ CO lines 
\citep{Storey1981, Watson1985, Jaffe1987, Boreiko1991}. 
Starting from the first PDR-models, tailored to explain the FIR line
emission from massive star forming regions such as Orion 
\citep{Tielens1985, Sternberg1989} the inclusion of
more details of the heating and cooling mechanisms as well as of the
chemical network, nowadays allows the modeling 
over a wide range of physical
parameters and has resulted in successful modeling of many detailed
aspects of observed photon-dominated regions 
\citep[review by][]{Hollenbach1999}. 
PDRs thus have proven to be
a very useful concept in understanding the mutual interaction between
star formation and the structure of the ISM,
both in individual regions in the Milky Way, but also in star forming
regions in external galaxies
\citep{Kramer2005, Contursi2002, Malhotra2001, Unger2000, Nikola1998, 
Madden1997b, Schilke1993}.
An overview of the different models is given by the recent comparison
study of PDR codes \citep{Roellig2007}.

The importance of the PDR scenario is related to the fact that
the ISM is fractal and thus most of the material is close to 
surfaces and hence affected by UV radiation, i.e. is located in
PDRs. This was realized early on by comparing the spatial
distribution of observed PDR tracers \citep{Stutzki1988, Howe1991} 
with simple models of homogeneous or clumpy cloud structures. 
The observed relatively uniform line ratios of low-$J$ \thirteenco{}
and \twelveco{} in molecular clouds, inconsistent with
simple, uniform cloud models, are shown to be
naturally explained if the emission is assumed to originate in many,
relatively small clumps \citep{Stoerzer1996}.
The necessity of high densities in order to explain the observed
large \thirteenco{} brightnesses in the mid-$J$ lines in the submm 
independently indicate a clumpy structure 
\citep{Graf1990, Wolfire1989}.

The fractal structure of the ISM is well represented 
by an ensemble of clumps with a power law clump mass 
distribution, and a power law clump \msr{} 
\citep{Stutzki1998}.
Power law mass spectra have been derived by decomposing
the observed 3D-datacubes of emission of \co~isotopologues
into clumps by various methods 
\citep{Stutzki1990, Kramer1998, Williams1993, Williams1994}. 
The identified clumps typically show also a power law \msr, 
although obtaining significant coverage over more then 1.5 
to 2 orders of magnitude in length scale is difficult; 
combining low and high angular resolution observations of 
the Polaris Flare, \citet{Heithausen1998} demonstrated a 
power law \msr{} over 3 orders of magnitude in length scale. 
Calculating the PDR emission from an ensemble of clumps with 
a given mass spectrum and \msr{} thus offers a convenient way 
to model the submillimeter and FIR line emission of the ISM 
as an UV penetrated clump ensemble with the given fractal 
characterstics.

The far-infrared absolute spectrophotometer (FIRAS) 
on the COBE satellite conducted a spectral line survey of 
the Milky Way in the far-infrared region at wavelengths 
longward of 100 \mum{} \citep{Wright1991, Bennett1994}. 
The spatial and spectral resolutions of respectively 
7\degr{} and $0.45\ \mathrm{cm}^{-1}$ allowed detection of
the CO rotational transitions from $J=1$--0 to 8--7, and 
the fine structure transitions of \cii~158 \mum, \nii~122 
and 205 \mum, \oi~146 \mum, and \ci~370 and 609 \mum, in 
addition to the dust FIR continuum. 
The wavelength coverage of COBE FIRAS did not allow to detect a
few other important cooling lines of the ISM like the \oi{} line at
63 \mum{} or the \oiii{} line at 88 \mum. 
\citet{Fixsen1999}
report the distribution of spectral line emission along the Galactic
plane. 
They re-binned the data to a $5\degr$-grid in Galactic longitude and
assumed an latitudinal extension of the FIR line emission of $1\degr$
according to the results of the DIRBE FIR continuum observations.
In a first attempt to interpret the observed emission with PDR
models, they fit the plane-parallel PDR model of Hollenbach (1991) to
the observed line ratios in the Galactic center region to derive a
density of $30\ \pcc$ and FUV field of 10 Habing units\footnote{ 
1 Habing unit corresponds to the integrated flux in the wavelength range
from 91.2 to 111.0 nm of $1.6 \times 10^{-3}$ erg \ps~\psc{} \citep{Habing1968}.
}.
\citet{Heiles1994} used the COBE Milky Way data to analyze the 
properties of the extended ionized gas traced by the \nii{} lines. 
\citet{Misiriotis2006} combined COBE DIRBE and FIRAS continuum data 
at wavelengths between
1 \mum{} and 1 mm wavelength to constrain exponential axisymmetric
models for the spatial distribution of the dust, the stars, and the
gas in the Milky Way.

In this paper we investigate to what degree the global FIR-line
emission distribution of the Milky Way as 
observed by COBE can be explained in the
framework of an UV-penetrated, clumpy cloud scenario of the ISM. 
We mention that the \ci\ 1--0 and 2--1 and \co\ 4--3 and 7--6 
emission from a massive star forming region in the Carina nebula 
can be explained by the emission of clumpy PDRs as recently shown 
by \citet{Kramer2008}.
The paper is organized as follows:
In Section 2 we shortly summarize the 
concepts to quantify the clumpy, fractal cloud structure 
and introduce the
clumpy cloud PDR model.
Section 3 discusses the line emissivity of the clump
ensemble, in particular the dependence on varying
mass limits of the clump distribution.
In Section 4 
the model results are compared with the observational data from COBE. 
A short conclusion with outlook is given in Section 5.


\section{The galaxy model: an UV-illuminated clump ensemble}
\label{s_model}

Modeling the large-scale Galactic FIR line emission as
observed by COBE is done in the following steps: 

\begin{itemize}
\item
generate a large parameter grid for the emission of individual
spherical clumps with the \kosma, where the individual clumps are
characterized by their mass, density (resp.\ size), and the
UV-intensity that they are embedded in. 
\item
generate a clump mass and size distribution at each Galactocentric
radius from the observed mass density and volume
filling factor distribution 
in the Milky Way, assuming an universal power law index for the clump mass and 
size spectra and applying adequate high and low mass cutoffs,
\item
estimate the Galactic radial distribution of the effective mean 
UV-field in which the dense ISM clumps are embedded 
and the average clump density in the ensemble respectively 
the volume filling factor of the clump ensemble,
\item
by combining the above, calculate the volume emissivity of the ISM 
originating from the UV illuminated clump ensemble as a function of 
Galactocentric radius in the Milky Way
\item
calculate the line emission versus Galactic longitude 
by line-of-sight integration over this emissivity in the direction towards
the Earth.
\end{itemize}
We describe each of these steps in detail in the following:

\subsection{
The KOSMA-$\tau$ PDR model: line intensities of 
an individual clump
}
\label{s_kosma}

\begin{figure}
\centering
\resizebox{\hsize}{!}{\includegraphics{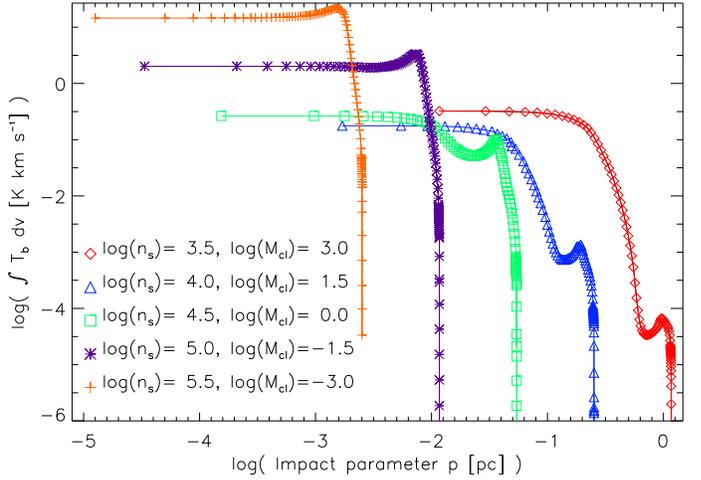}}
\caption{Brightness temperature 
 vs. impact parameter $p$ of the \co~7--6 rotational transition for 
 selected clumps with mass and density values following
 the \cms{} and \msr{}. $n_{\ind{s}}$ is the gas density at
 the surface of the clumps in [\pcc] (see Eq. \ref{e_ncl}), 
 $M_{\ind{cl}}$ the clump mass in [\msun], and the incident 
 FUV-flux is $\chi = 10^2\ \fdraine$.
}
\label{f_CO_7-6_clumpint}
\end{figure}

\begin{table*}
\caption{
Properties of the \kosma~clumps with given masses and densities at an
incident FUV-flux of $10^2$ \fdraine.
Orders of magnitude are given in parentheses.
}
\label{t_kosma_info}
\centering
\begin{tabular}{ccccccc}
\hline \hline
\multicolumn{2}{c}{Densities} & Mass & \multicolumn{2}{c}{Column densities} & 
\multicolumn{2}{c}{Temperatures} \\
Surface & Center & & \multicolumn{4}{c}{Clump average} \\
$\log(n\ind{s}/\pcc)$ & $\log(n\ind{c}/\pcc)$ & $\log(M\ind{cl}/\msun)$ & 
$\left< N_{\htwo} \right>$ & $\left< N_{\co} \right>$ & 
$\left< T_{\htwo} \right>$ & $\left< T_{\co} \right>$ \\
\hline
3.5 & 4.55 & 3.0 & 1.41(22) & 1.71(18) & 21.7 & 12.3 \\
4.0 & 5.05 & 1.5 & 9.74(21) & 1.50(18) & 23.8 & 11.0 \\
4.5 & 5.55 & 0.0 & 6.69(21) & 1.07(18) & 26.5 & 12.5 \\
5.0 & 6.05 & -1.5 & 4.57(21) & 7.21(17) & 29.6 & 15.5 \\
5.5 & 6.55 & -3.0 & 3.12(21) & 4.94(17) & 32.8 & 19.9 \\
\hline
\end{tabular}
\end{table*}

The \kosma{} describes spherical clumps characterized by their mass, 
density (or size, respectively), and the incident UV field. 

We treat the mass of the clumps $M\ind{cl}$ as the total 
hydrogen gas mass.
To obtain the total gas mass of a clump the mass of helium and
heavier elements has to be added. This is a factor of about 1.4
higher \citep{Anders1989}.
Due to the fact, that the massive clumps dominate the ensemble mass
(see Section \ref{s_clump_dist})  
and contain only a few percent of their gas mass in the atomic 
phase, we can neglect the contribution of the atomic gas
so that the ensemble mass is basically given by the mass of
molecular hydrogen.

The radial density structure $n(r)$ of the 
model clumps with radius $R\ind{cl}$ is given by
an inner core region with constant density for $r < 0.2\,R\ind{cl}$ and a
power law with index $-1.5$ for $0.2\,R\ind{cl} < r < R\ind{cl}$:

\begin{equation}
n(r) = n_{\mathrm s} \cdot
 \begin{cases}
  (\frac{r}{R\ind{cl}})^{-1.5} & \text{, for } 0.2\,R\ind{cl} < r < R\ind{cl} \\
  5^{1.5} & \text{, for } r \le 0.2\,R\ind{cl} \\ 
 \end{cases}
\label{e_ncl}
\end{equation}

Note that $ n\ind{s} = n\ind{\element{H},s} + 2 \, n\ind{\htwo,s} $ is the 
total hydrogen nucleus number density at the clump surface.
By neglecting the presence of helium and metals in the number density, 
we underestimate the number of collision partners by a factor of $1.2$. 
The effect on the collision rates is, however, smaller due to
the larger mass and thus lower velocity of helium.
This does not change the results significantly.
With the clump mass $M\ind{cl}$, the
clump volume $V\ind{cl} = \frac{4\pi}{3}\,
R\ind{cl}^3$, and the 
mass of the hydrogen atom m\ind{\element{H}},
the average clump density $ n\ind{cl} = 
M\ind{cl} / (\mathrm{m}_{\element{H}} V\ind{cl}) $,
used in the following to characterize each clump, is given by 
$n\ind{cl} = \left( 2-5^{-3/2} \right) n\ind{s} \approx 1.91 \, n\ind{s}$.

The FUV flux
illuminating each clump is assumed to be isotropic. 
It is given in units of the Draine field 
$\fdraine = 2.7 \times 10^{-3}\ \ergpspsc$.
This is the flux of the local interstellar radiation 
field (ISRF) integrated over the wavelength interval 
from 91.2 to 111.0 nm \citep{Draine1978}.

In the \kosma~ a constant flux of cosmic rays and the 
incident FUV radiation are the only heating sources of 
the clumps. 
The temperature and chemical structure of the clumps are 
derived from a radiative transfer computation including 
line-shielding for the incident FUV radiation and an 
escape probability approximation for the emitted FIR 
line radiation.
The efficiency of the line cooling depends on the average 
line width $\delta v$, which is assumed to be 1.2 km \ps{} 
(see Table \ref{t_kosma_par}). 
More details are described by \citet{Gierens1992}, 
\citet{Koester1994}, \citet{Stoerzer1996}, and 
\citet{Roellig2006}.

In order to calculate the average line intensity 
$I\ind{cl}$ of each clump, first a line-of-sight 
integration of the source function and the attenuation 
along parallel paths through the clump is performed
to calculate the resulting intensities $I(p)$ for 
different impact parameters $p$.
As an example the line integrated \co~7--6 intensities
are shown as a function of impact parameter in Fig. 
\ref{f_CO_7-6_clumpint}.
Note the strong emission of the small, low mass, 
high density clumps.
The physical reason for this is the enhanced 
population of the $J=7$ rotational excited state 
(i.e. a higher excitation temperature) of the 
\co~molecule in the smaller and denser clumps, 
as only for those the density in the clump center 
falls near or above the critical density of the 
\co~7--6 rotational transition of about 
$10^6\ \pcc$ \citep{Kaufman1999, Kramer2004}.
The less dense but massive clumps have much larger
column densities, partially compensating for the lower
excitation temperature and increasing the optical
depth. 
This leads to the apparently constant central brightness
temperatures of the bigger clumps in Fig. 
\ref{f_CO_7-6_clumpint}.
Detailed information about these clumps is compiled in 
Table \ref{t_kosma_info}.

Second, $I\ind{cl}$ is obtained by averaging over the 
projected clump surface via

\begin{equation}
I\ind{cl}=\frac{2}{R\ind{cl}^2} \int_0^{R\ind{cl}} 
I(p) \, p \, \mathrm{d}p.
\label{e_I_cl}
\end{equation}

The results of these computations are used for the clump 
ensemble intensities. 

We generate a large grid of intensities for the three 
independent clump parameters mass, density and 
FUV intensity, with grid points equidistant 
on a logarithmic scale in steps of half an order of 
magnitude.
The range covered by the parameter cube of the \kosma{}
is shown in Table \ref{t_kosma_par_cube}.
For clumps with intermediate parameter values,
the \kosma{} intensities are interpolated between the 
grid points using logarithmic differences. 
Further parameters for the PDR model, held fixed for 
the present work, are listed in Table \ref{t_kosma_par}.

\begin{table}
\caption{The parameter values of the precomputed \kosma{}
 clumps are equidistant on logarithmic scale in steps of 
 half an order of magnitude.
} 
\label{t_kosma_par_cube}
\centering
\begin{tabular}{lc}
\hline \hline
Quantity & Parameter range \\
\hline
Density $n\ind{s} / \pcc$ & $10^2$, \ldots, $10^6$ \\
Mass $M\ind{cl} / \msun$ & $10^{-3}$, \ldots, $10^3$ \\
FUV flux $\chi / {\fdraine}$ & $10^0$, \ldots, $10^6$ \\
\hline
\end{tabular}
\end{table}

\begin{table}
\caption{
The fixed parameter values of the precomputed {\kosma}~clumps. Note that
powers of ten are given in parentheses
and the elemental abundances are given in units of the hydrogen abundance,
i.e. $ X_{i} = n_{i} / n $.
}
\label{t_kosma_par}
\centering
\begin{tabular}{lr}
\hline \hline
Quantity & Value \\
\hline
\element{He} abundance $X_{\element{He}}$ & 0.1\\
\element{O} abundance $X_{\element{O}}$ & $3.0 \, (-4)$ \\
\element{C} abundance $X_{\element{C}}$ & $1.4 \, (-4)$ \\
\element{S} abundance $X_{\element{S}}$ & $2.8 \, (-5)$ \\
\element[][13]{C} abundance $X_{\element[][13]{C}}$ & $2.1 \, (-6)$ \\
Cosmic ray ionization rate $\chi_{\mathrm CR}\ [\ps] $ & $5 \, (-17)$ \\
Velocity width (FWHM) $ \delta v\ [\mathrm{km}\ \ps] $ & 1.2 \\
Dust UV cross section $\sigma\ind{dust}\ [\mathrm{cm}^2]$ & $1.9 \, (-21)$ \\
\hline
\end{tabular}
\end{table}

\subsection{Clump mass distribution and geometry}
\label{s_clump_dist}

Observations of the ISM show structure at all scales down to the 
resolution limit, as studied by e.g. \citet{Stutzki1988},
\citet{Stutzki1990}, \citet{Heithausen1998}, \citet{Kramer1998}, 
\citet{Simon2001}.
Quantitative analysis of the observed intensity distribution shows that it
can be described as a self-similar, i.e. fractal structure.
Alternatively, the observed 3d spectral line data cubes can be decomposed into
clump ensembles.
The ``\cms'', i.e.\ the number of clumps $\diff{N\ind{cl}}$ per mass bin
$\diff{M\ind{cl}}$, is then given by
\begin{equation}
\ddiff{N\ind{cl}}{M\ind{cl}} = A \, M\ind{cl} ^ {-\alpha}.
\label{e_cms}
\end{equation}
The size of the clumps is correlated with their mass via the ``\msr'':
\begin{equation}
M\ind{cl}  = C \,  R\ind{cl} ^\gamma .
\label{e_msr}
\end{equation}
The indices $\alpha$ and $\gamma$ of the clump ensemble 
are related to the fractional Brownian-motion power law 
index characterizing the fractal structure of the ISM 
\citep{Stutzki1998}. 
We choose values of 1.8 and 2.3, respectively,
following the results from \citet{Heithausen1998}.
The ensemble mass $M\ind{ens}$ is obtained by integrating 
the \cms~over a mass range from a low mass cutoff $m\ind{l}$ 
to a high mass cutoff $m\ind{u}$.
The total mass of the clump ensemble 
$ M\ind{ens} = \int_{m\ind{l}}^{m\ind{u}} M\ind{cl} \,
\ddiff{N\ind{cl}}{M\ind{cl}} \, \diff{M\ind{cl}} $
determines the constant of proportionality $A$ in 
Eq. \ref{e_cms} via
\begin{equation}
M\ind{ens} = \frac{A}{2-\alpha} 
\left( m\ind{u}^{2-\alpha}-m\ind{l}^{2-\alpha} \right).
\end{equation} 
Note that for $1 < \alpha < 2$
the bulk of the ensemble mass is contained in the 
high mass clumps. 
In contrast, the total number of clumps 
$ N\ind{ens} = \int_{m\ind{l}}^{m\ind{u}} 
 \ddiff{N\ind{cl}}{M\ind{cl}} \, \diff{M\ind{cl}}
= \frac{A}{1-\alpha} 
 \left( m\ind{u}^{1-\alpha}-m\ind{l}^{1-\alpha} \right)$
is dominated by the low mass clumps. 
Using the volume $V\ind{cl} = \frac{4\pi}{3} R^3\ind{cl}$
of a clump of mass $M\ind{cl}$ and the \msr, 
the volume of the clump ensemble 
$ V\ind{ens} = \int_{m\ind{l}}^{m\ind{u}} 
\ddiff{N\ind{cl}}{M\ind{cl}} \, V\ind{cl} \, 
\diff{M\ind{cl}} $ is given by
\begin{equation}
V\ind{ens} = \frac{A}{C^{3/\gamma}} \frac{4\pi}{3}
 \frac{m\ind{u}^{1+3/\gamma-\alpha} - 
 m\ind{l}^{1+3/\gamma-\alpha}}{1+3/\gamma-\alpha}
\label{e_V_ens}
\end{equation}
For the values of $\alpha$ and $\gamma$ given above, the exponent is
$1+3/\gamma-\alpha \approx 0.5$, and hence the ensemble volume is
dominated by the large and most massive clumps.

The above applies as long as we are concerned with the integral
properties of the clump distribution. In order to take the spatial
distribution of the clumpy medium into account, we now switch to a volume
specific notation where the medium is characterized by the (volume) number
density, rather than the number, of the clumps, and by the mass density rather
than the total mass. 
We obtain for the volume- and mass-differential clump distribution, i.e. for the
(volume) number density of clumps $\diff{\tilde N\ind{cl}}$ per clump mass interval
$\diff{M\ind{cl}}$, the ``clump number density spectrum''
as a function of Galactocentric radius $\vec{r}$:
\begin{equation}
\ddiff{\tilde N\ind{cl}(\vec{r})}{M\ind{cl}} =
 \tilde A(\vec{r}) M\ind{cl}^{-\alpha}.
\end{equation}
The ``clump mass density spectrum'', i.e. the mass per volume element, 
per mass interval of the clump ensemble is correspondingly given by
\begin{equation} 
\ddiff{\tilde N\ind{cl}(\vec{r})}{M\ind{cl}}
 \, M\ind{cl} =
 \tilde A(\vec{r}) M\ind{cl}^{1-\alpha},
\end{equation}
and the mass differential clump volume per volume element, 
i.e. the ``clump filling factor spectrum'', is
\begin{equation}
\ddiff{\tilde N\ind{cl}(\vec{r})}{M\ind{cl}}
 \, V\ind{cl} =
 \tilde A(\vec{r}) C^{-3/\gamma}(\vec{r})  
 M\ind{cl}^{3/\gamma-\alpha}.
\end{equation}
Integration over the clump mass leads to
the ``volume mass density'' $\rho(\vec{r})$ 
and the ``volume filling factor'' $f_V(\vec{r})$,
respectively, given by
\begin{eqnarray}
\rho(\vec{r}) &=& 
\frac{ \tilde A(\vec{r}) }{2-\alpha} 
\left( m\ind{u}^{2-\alpha}-m\ind{l}^{2-\alpha} \right) 
{\rm~,~and} 
\label{e_rho} 
\\
f_V(\vec{r}) &=&  \frac{4\pi}{3} 
 \frac{ \tilde A(\vec{r}) }{ C^{3/\gamma}(\vec{r}) }
 \frac{m\ind{u}^{1+3/\gamma-\alpha} - m\ind{l}^{1+3/\gamma-\alpha}}
 {1+3/\gamma-\alpha}.
\label{e_f_V}
\end{eqnarray}
The ``ensemble averaged clump mass density''
$\rho\ind{ens}(\vec{r}) =
\rho(\vec{r}) / f_V(\vec{r})$ is given by
\begin{eqnarray}
\rho\ind{ens}(\vec{r}) &=& 
\frac{3}{4\pi}
\, C^{3/\gamma}(\vec{r})\,
\frac{1+3/\gamma-\alpha}{2-\alpha} \notag \\
&\cdot&
\frac{ (m\ind{u}^{2-\alpha} - m\ind{l}^{2-\alpha}) }
{ (m\ind{u}^{1+3/\gamma-\alpha} - m\ind{l}^{1+3/\gamma-\alpha}) }
\label{e_rho_ens}
\end{eqnarray}

As discussed in detail in subsection 2.4 below, the distribution of the dense
ISM in the Milky Way in our model 
is specified by the volume mass density $\rho(\vec{r})$
and the (higher) ensemble averaged clump mass density 
$\rho\ind{ens}(\vec{r})$ 
(respectively the volume filling factor of the molecular material). 
For given values of the low and high mass cutoffs of the clump distribution,
equations (\ref{e_rho}) and (\ref{e_rho_ens}) then determine
the pre-factor
of the clump number density spectrum  and its spatial variation,
\begin{equation}
\tilde A(\vec{r})=
\frac{2-\alpha}
{m\ind{u}^{2-\alpha}-m\ind{l}^{2-\alpha}} \, \rho(\vec{r}),
\label{e_A}
\end{equation}
through the volume mass density profile
of the Milky Way, and the pre-factor 
of the \msr{} and its spatial variation,
\begin{eqnarray}
&& C(\vec{r}) = \notag \\
&& \left( \frac{4 \pi}{3}\,
\frac{2-\alpha}{1+3/\gamma - \alpha}\,
\frac{m\ind{u}^{1+3/\gamma -\alpha}-m\ind{l}^{1+3/\gamma -\alpha}}
    {m\ind{u}^{2-\alpha}-m\ind{l}^{2-\alpha}}\,
\rho\ind{ens}(\vec{r})
\right)^{\gamma/3}
\label{e_C}
\end{eqnarray}
through the ensemble averaged clump mass density.

A further quantity of interest is the
beam filling factor $f_{\rm B}$, 
as the ratio of the
solid angle filled by the clumps of the ensemble 
$\Omega\ind{ens}$ 
and the solid angle of the beam $\Omega_{\rm B}$.
The solid angle of a specific clump with mass $M\ind{cl}$ and
Radius $R\ind{cl}$ at a distance $s=|\vec s|$ 
(considering $\vec s$ as given in Galactic
coordinates centered at the sun) is given by
\begin{equation}
 \Omega\ind{cl}(s) 
= 2 \pi (1-\sqrt{1-(R\ind{cl} / s)^2}) 
\simeq \pi \frac{R\ind{cl}^2}{s^2}. 
\end{equation}
for $R\ind{cl} \ll s$.
Then, the ``solid angle density spectrum'',
i.e. the solid angle
per volume and clump mass element
of the ensemble, is
\begin{equation}
 \omega\ind{cl}(\vec{s}) =
\frac{d\tilde N\ind{cl} }{ dM\ind{cl} } \, 
\Omega\ind{cl} 
\simeq \pi 
\frac{ \tilde A(\vec{s}) }{ C^{2/\gamma}(\vec{s}) 
\, s^2 } \, M\ind{cl}^{2/\gamma - \alpha}.
\end{equation}
For the adopted values of 
$\alpha$ and $\gamma$ the exponent $2/\gamma - \alpha$ 
has a value of $-0.93$, implying that 
on a logarithmic mass scale,
$ \omega\ind{cl}(\vec{s}) \, \diff{M\ind{cl}}/\diff{(\ln M\ind{cl})}
= M\ind{cl} \, \omega\ind{cl}(\vec{s}) 
\propto M\ind{cl}^{0.07} $,
each logarithmic mass interval contributes about equal to the solid angle of
the ensemble.
Integration over the given clump mass range
gives the ``solid angle density'' of the 
ensemble
\begin{eqnarray}
&& \omega\ind{ens}(\vec s) = 
\int_{m\ind{l}}^{m\ind{u}} \omega\ind{cl}(\vec s) \, \diff{M\ind{cl}} \notag \\
&& \simeq \pi \frac{ \tilde A(\vec{s}) }
{ C^{2/\gamma}(\vec{s}) \, s^2 } \, 
\frac{ m\ind{u}^{2/\gamma - \alpha + 1} - m\ind{l}^{2/\gamma - \alpha + 1} }
{2/\gamma - \alpha + 1}  \notag \\
&& = \pi^{1/3} \left( \frac{3}{4} \right)^{2/3} \,
\frac{\rho(\vec{s})}{\rho\ind{ens}(\vec{s})^{2/3}\, s^2} \,
\frac{(2-\alpha)^{1/3}(1+3/\gamma -\alpha)^{2/3}}
     {2/\gamma - \alpha +1 } \notag \\
&& \cdot
\frac{m\ind{u}^{2/\gamma -\alpha +1}-m\ind{l}^{2/\gamma -\alpha +1}}
     {(m\ind{u}^{2-\alpha}-m\ind{l}^{2-\alpha})^{1/3}
      (m\ind{u}^{1+3/\gamma -\alpha}-m\ind{l}^{1+3/\gamma -\alpha})^{2/3}}.
\end{eqnarray}
For the values of $\alpha$ and $\gamma$ adopted above, 
the nominator shows again a very weak mass dependence of $\propto m^{0.07}$. 
In the denominator in both factors the upper-mass cut-off dominates, so that
the total dependence of the solid angle volume density of the clump ensemble
is $\propto m\ind{u}^{-(4/3-\alpha +2/\gamma)} \simeq m\ind{u}^{-0.4}$, i.e.\
weakly decreasing with increasing upper mass cut-off and basically
independent on the lower mass cut-off. 
Integration over the beam volume $V_{\rm B}$
gives the total solid angle filled by the clump ensemble 
\begin{eqnarray}
&& \Omega\ind{ens} =  \int_{V_{\rm B}} 
\omega\ind{ens}(\vec s) \, \diff{V} \notag \\
&& \simeq 
\pi^{1/3} \left( \frac{3}{4} \right)^{2/3} \,
\int_{V_{\rm B}} 
\frac{\rho(\vec{s})}{\rho\ind{ens}(\vec{s})^{2/3}\, s^2} \diff{V} \notag \\
&& \cdot  
\frac{(2-\alpha)^{1/3}(1+3/\gamma -\alpha)^{2/3}}
     {2/\gamma - \alpha +1 } \notag \\
&& \cdot  
\frac{m\ind{u}^{2/\gamma -\alpha +1}-m\ind{l}^{2/\gamma -\alpha +1}}
     {(m\ind{u}^{2-\alpha}-m\ind{l}^{2-\alpha})^{1/3}
      (m\ind{u}^{1+3/\gamma -\alpha}-m\ind{l}^{1+3/\gamma -\alpha})^{2/3}}.
\end{eqnarray}
with the same dependence on upper and lower mass cut-off as for the volume
specific solid angle. 


\subsection{Volume emissivity and intensity of the clumpy medium}
\label{s_emint}

In order to calculate the intensity of the clump ensemble, we neglect any
absorption (in particular self-absorption in the spectral lines) 
among different clumps along the line-of-sight. 
This assumption may be violated locally 
in the high densities of individual star forming regions,
although even there, the higher velocity dispersion of the virialized clump
ensemble helps to avoid line-of-sight crowding in each velocity interval. 
It is reasonable for the large scale FIR line emission 
of the Galaxy, taking the additional velocity spread between
individual regions due to the differential Galactic rotation 
into account.
The intensity then simply is given by the 
line-of-sight integration over the volume emissivity of the clump ensemble
\begin{equation}
 I = \int\ind{l} \eta(s) \, \diff{s}
\end{equation}  
The beam average intensity then is
\begin{equation}
  I\ind{B} = 
\frac{1}{\Omega\ind{B}} \int_{\Omega\ind{B}} \int\ind{l} \eta(\vec{s}) \, \diff{s} \, \diff{\Omega}
= \frac{1}{\Omega\ind{B}} \int_{V\ind{B}} \eta(\vec{s})\, \frac{\diff{V}}{s^2}.
\label{e_I_B}
\end{equation} 
The \kosma{} gives the clump average specific intensity $I\ind{cl}$ for
each clump of a given mass. The ``volume emissivity spectrum''
then is given
by multiplication with the clump projected area $ A\ind{cl} = s^2 \,
\Omega\ind{cl} $ and the clump number density spectrum, namely
\begin{equation}
\ddiff{\eta(\vec{s})}{M\ind{cl}} = \ddiff{\tilde N\ind{cl}}{M\ind{cl}} 
\, s^2 \, \Omega\ind{cl} \, I\ind{cl}. 
\end{equation}
With $ \omega\ind{cl} = 
\ddiff{\tilde N\ind{cl}}{M\ind{cl}}\,\Omega\ind{cl} $, 
we obtain for the contribution to the ensemble intensity in the mass bin
$\diff{M\ind{cl}}$
\begin{equation}
\ddiff{I\ind{B}}{M\ind{cl}} = 
\frac{1}{\Omega\ind{B}} \int_{V\ind{B}} I\ind{cl} \, \omega\ind{cl}(\vec{s}) \, \diff{V}.
\end{equation}
Integration over the mass ensemble
results in the beam averaged intensity of the ensemble
\begin{equation}
I\ind{B} = \frac{1}{\Omega\ind{B}} 
\int_{V\ind{B}} \int_{m\ind{l}}^{m\ind{u}} 
I\ind{cl} \, \omega\ind{cl}(\vec{s}) \, \diff{M\ind{cl}} \, \diff{V}.
\label{e_I_B_ens}
\end{equation} 
Particularly, we obtain for a constant clump intensity
the resulting beam average intensity $I\ind{B}$ 
with the beam filling factor 
$f\ind{B} = \Omega\ind{ens} / \Omega\ind{B}$
as $I\ind{B} = f\ind{B} \, I\ind{cl}$.

\subsection{Galactic parameter distributions}
\label{s_galpar_dist}

\begin{table*}
\caption{
 Parameters of the Galaxy model (values at the solar circle). --
 * For distribution over Galactocentric radius see Fig. \ref{f_galpar} or text. --
 ** Assumed to be constant over the whole Galactic disk (see text). --
 [1] \cite{Wolfire2003} --
 [2] \cite{Reynolds1991} --
 [3] \cite{Heithausen1998}
}
\label{t_galpar}
\centering
\begin{tabular}{lllrl}
\hline \hline
Name & Symbol & Unit & Value & Reference\\
\hline
Radial extension of Galactic disk & R$\ind{G}$ & kpc & 18 & [1]\\
Solar distance to Galactic center & R$_{\sun}$ & kpc & 8.5 & [1]\\
Half height H$_2$ ** & h$\ind{\htwo}$ & pc & 59 & [1]\\
Half height CNM/WNM ** & h$\ind{\hi}$ & pc & 115 & [1]\\
Half height WIM ** & h$\ind{WIM}$ & pc & $10^3$ & [2] \\
H$_2$ mass surface density * & $\Sigma\ind{\htwo}$ & \msun{} pc$^{-2}$ & 1.4 & [1] \\
CNM mass surface density * & $\Sigma\ind{CNM}$ & \msun{} pc$^{-2}$ & 2.25 & [1] \\
WNM mass surface density * & $\Sigma\ind{WNM}$ & \msun{} pc$^{-2}$ & 2.75 & [1] \\
WIM mass surface density * & $\Sigma\ind{WIM}$ & \msun{} pc$^{-2}$ & 1.85 & [2,1] \\
Ensemble averaged clump density * & $n\ind{ens}$ & \pcc{} & $10^{3.8}$ & \\
CNM density * & $n\ind{CNM}$ & \pcc{} & 32.9 & [1] \\
WNM/WIM density * & $n\ind{WNM}$ & \pcc{} & 0.349 & [1] \\
CNM temperature * & $T\ind{CNM}$ & K & 85 & [1] \\
WNM/WIM temperature * & $T\ind{WNM}$ & K & 7860 & [1] \\
FUV-flux ** & $\chi$ & \fdraine & $10^{1.8}$ & \\
Metallicity ** & $Z$ & Z$_{\sun}$ & 1 & \\
Clump mass spectral index ** & $\alpha$ & & 1.8 & [3] \\
Mass-size relation index ** & $\gamma$ & & 2.3 & [3] \\
Upper clump mass limit ** & $m\ind{u}$ & \msun{} & $10^{2}$ & [3] \\
Lower clump mass limit ** & $m\ind{l}$ & \msun{} & $10^{-3}$ & [3] \\
\hline
\end{tabular}
\end{table*}

\begin{figure}
\centering
\resizebox{8cm}{!}{\includegraphics{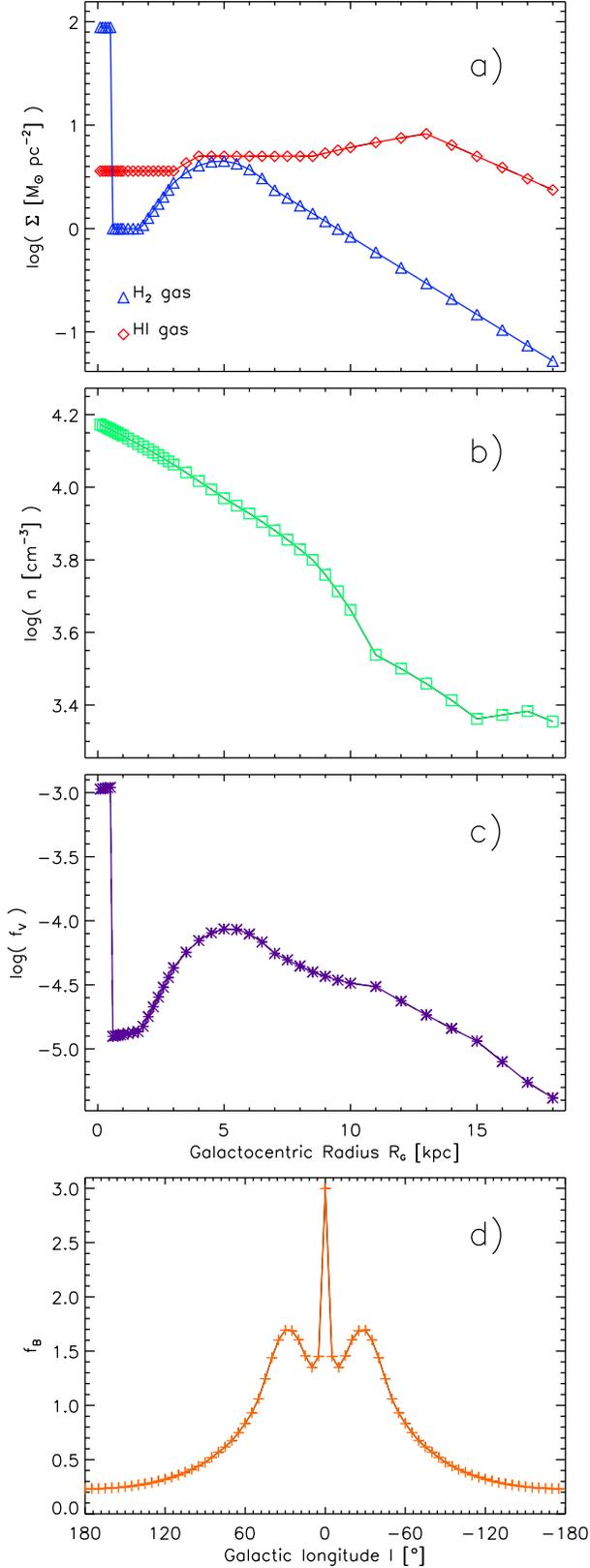}}
\caption{
 {\textbf a)} Galactic mass surface density distributions 
 of the molecular and atomic gas mass;
 {\textbf b)} ensemble averaged clump density 
 distribution;
 {\textbf c)} resulting volume filling factor distribution
 with Galactocentric radius;
 {\textbf d)} derived beam filling factor distribution
 with Galactic longitude from the solar point of view.
}
\label{f_galpar}
\end{figure}

The Galactic parameter distributions of mass, density, and UV-intensity
are assumed to vary only in Galactocentric radial direction within
a cylindrically symmetric Galactic disk of constant half-height $h$.
The values of the different parameters 
(at the solar circle at R = R$_\odot=8.5$ kpc)
are compiled in Table \ref{t_galpar}.
We adopt a half-height of the molecular disk of
59 pc and for the atomic gas layer of 115 pc 
\citep{Wolfire2003}.
The fractal characteristics ($\alpha$, $\gamma$, $m\ind{l}$, $m\ind{u}$)
of the clump distribution are assumed to be 
constant over the whole Galaxy.
Due to the simplified structure of the Galactic parameter
distributions in our model disk we do not expect to
reproduce individual emission features along the Galactic 
plane.
The Galactic mass distribution of the neutral ISM is a key 
parameter of the model, because the modeled line intensities are 
basically proportional to the mass of the molecular gas assumed to 
be completely represented by PDRs (see Section \ref{s_intro}).
The Galactic mass distribution has been discussed in the literature 
e.g. by \citet{Clemens1985}, \citet{Rohlfs1987}, 
\citet{Williams1997}, \citet{Bronfman2000}.
We use an approximation of the radial Galactic H$_2$ 
mass distribution presented by \citet{Wolfire2003} for 
3 $<R_G/\text{kpc}<$ 18.
According to \citet[Fig. 3]{Williams1997} the H$_2$
mass surface density decreases inwards from 3 to 1.7 kpc
Galactocentric radius exponentially by a factor of about
3/8.
For 0.6 $<R_G/\text{kpc}<$ 1.7 the molecular hydrogen gas
mass distribution is set constant.
In the Galactic center region ($R_G<0.6$ kpc), the molecular 
hydrogen gas mass surface density is assumed to have a constant 
value of 88.4 \msun~pc$^{-2}$ corresponding to an 
H$_2$ mass of $10^8$ \msun~within the central 600 pc 
of the Galaxy \citep{Dahmen1998, Guesten2004}.
This translates into a mean H$_2$ gas mass density of 
0.75 \msun~pc$^{-3}$, corresponding to an H$_2$ number 
density of about 15 cm$^{-3}$.
The assumed mass distributions are visualized in Fig. 
\ref{f_galpar} a).
The neutral atomic hydrogen mass distribution 
is used to estimate the contribution
of diffuse atomic phases to the \cii~158\mum{}
fine structure line emission (as shown in Section
\ref{s_cii-discussion}).
The contribution of the diffuse WIM is estimated 
by \citet{Reynolds1990} to 37\% of the total \hi\ mass,
which is slightly less than the WNM mass.

We adopt a lower clump 
mass limit $m\ind{l}$ of $10^{-3}$ \msun{} according to the 
observational lower clump mass limit from \citet{Heithausen1998},
as well as an upper clump mass limit $m\ind{u}$ of 
$10^{2}$ \msun{}.
\citet{Gorti2002} show an decreasing lifetime of clumps 
with decreasing mass and increasing density 
and argue against the existence of stable,
small, and dense PDR clumps.
This indicates a transient nature of these clumps. 

The observed mean gas density of molecular clouds
shows variation with Galactocentric radius 
\citep[e.g.][]{Brand1995}. 
We adopt the Galactic radial distribution of the CNM density 
provided by \citet{Wolfire2003} for the 
ensemble averaged clump density 
$ n\ind{ens} = \rho\ind{ens} / \mathrm{m}_{\element{H}} $,
scaled to an absolute value of 
$10^{3.8}\ \pcc$ at the solar circle 
(Fig. \ref{f_galpar} b)),
obtained by a $\chi^2$ fit to the intensity distributions 
as observed by COBE
(see section \ref{s_quantitative_comparison}).
As shown in Section \ref{s_clump_dist}, 
the volume filling factor is determined by 
the underlying mass and density distributions. 
The result is shown in Fig. \ref{f_galpar} c).
The computed beam filling factors 
from the solar point of view along the Galactic plane 
are shown in Fig. \ref{f_galpar} d).
The density and temperature distributions 
with Galactocentric radius of the CNM and WNM, 
used in the following to estimate the \cii{} emission
from these phases, are adopted from
\cite{Wolfire2003} (extrapolated inwards of 3 kpc
Galactocentric radius) and the WIM distributions
are assumed to equal the WNM distributions
\citep{McKee_Ostriker1977, Reynolds1991, Cox2005,
Hill2007}.

The FUV flux in the Galaxy is mainly produced by OB stars.
Observations show, that OB associations are associated
with giant molecular clouds (GMCs)
\citep[e.g.][]{Stark1978}.
\citet{McKee1997} and \citet{Williams1997} show,
that the distributions of Galactic OB associations
and GMCs can be described by 
truncated power laws with the same slope, i.e.
show a strong correlation.
\citet{Misiriotis2006} derive a nearly constant dust temperature
from the COBE observations. 
Assuming a proportionality between dust continuum
emission tracing the molecular material 
and the incident FUV flux from the stellar population this 
leads to a rather constant average FUV-flux impinging on the
molecular gas within the Galactic disk.
According to \citet[and references therein]{McKee1997}
there are 6-7 OB associations with
more than 30 massive stars within 
1 kpc distance from the sun, dominating the
local interstellar FUV radiation field. 
\citet{Blaauw1985} states a mean scale height of about
50 pc of the local OB associations.
Therefore we assume one layer of surrounding OB 
associations with 3 next neighbors.
This leads to an average radius of major influence
for each OB association of 
$R_{\rm OB} \approx 7^{-1/2}$ to $6^{-1/2}$ kpc or 
378 to 408 pc. 
The actual distance measurement to the Orion Nebula
Cluster resulting in $414 \pm 8$ pc \citep{Menten2007}
corresponds well to this range.
Together with the local interstellar FUV-flux
of 1 \fdraine{} \citep{Draine1978} and
average distances $d_{\rm MC-OB}$ of about 20 to 50 pc of the
GMCs to their associated OB associations
\citep[estimated from the sizes and shapes given in]
 [Fig. 1 and Table 1]{Stark1978} 
we obtain FUV-fluxes impinging the molecular
clouds of 
$\chi_{\rm MC} \approx (R_{\rm OB}/d_{\rm MC-OB})^2 \chi_{\rm D} / 3$
with values of 19 to 139 \fdraine{} or 
$\log( \chi_{\rm MC} / \chi_{\rm D}) \approx 1.3$ to $2.1$.
In Section \ref{s_results} we find a best fitting value
of $\log( \chi / \chi_{\rm D}) \approx 1.8$ for
our model, falling well within this range.


\section{
The clump-ensemble emissivity
}
\label{s_emissivity}

\begin{figure}
\centering
\resizebox{\hsize}{!}{\includegraphics{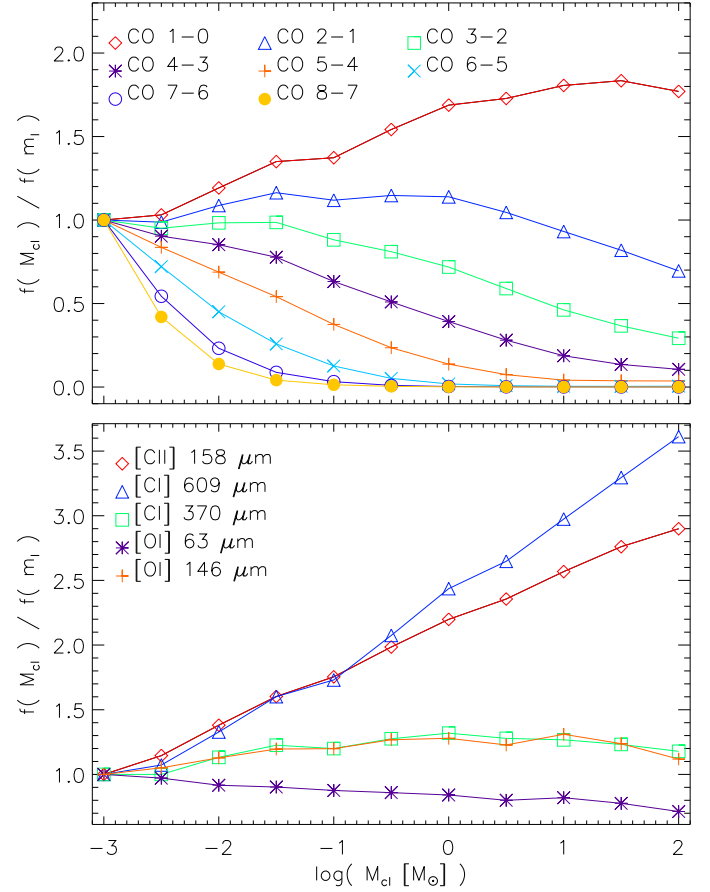}}
\caption{ 
 Volume emissivity spectra of the clump ensemble for an ensemble 
 averaged gas density of $n\ind{ens} = 10^{3.8}\ \pcc$ and an
 incident FUV-flux of $\chi = 10^{1.8}\ \fdraine$. 
 According to the logarithmic scaling of the clump-mass axis 
 $ f(M\ind{cl}) = M\ind{cl} \, \diff{\eta} / \diff{ M\ind{cl} }$ 
 is plotted to see the integral contribution of each mass interval.
 The values are normalized to the functional value at 
 $M\ind{cl} = m\ind{l}$ for a convenient comparison of the different
 tracers.
}
\label{f_mass-specific_emissivities}
\end{figure}

To discuss and understand the behavior of the
clump ensemble emissivity it is useful to
first look at the features of individual clumps 
and the mass-differential emissivity of the ensemble.
As explained in Section \ref{s_emint} the 
contribution to the volume emissivity spectrum
of the ensemble $\diff{\eta}/\diff{M\ind{cl}}$
in the various line transitions is dominated by the 
clump intensity $I_{\rm cl}$, as the filling
factors do not significantly depend on the 
clump mass.
The logarithmic volume emissivity spectrum
$\diff{\eta}/\diff{\log(M\ind{cl})} = 
M\ind{cl}\,\diff{\eta}/\diff{M\ind{cl}}$
for an ensemble averaged clump density of 
$n\ind{ens} = 10^{3.8}\ \pcc$ and an FUV-flux of
$\chi = 10^{1.8}\ \fdraine$
is shown in Fig. \ref{f_mass-specific_emissivities}.
The symbols show the sampling points on the model grid.
The upper plot shows the behavior for the different 
\co{} transitions. Note the trend towards 
larger contributions of the lower clump masses
for higher transition numbers.
The ensemble emission of the higher
\co{} transitions ($J>5$) almost exclusively 
stem from the low mass clumps.
The bottom panel shows
$M\ind{cl}\,\diff{\eta}/\diff{M\ind{cl}}$
for the fine structure cooling lines.
The \cii~158 \mum~and \ci~609 \mum~emissions
show an increased contribution of high mass clumps.

\begin{figure}
\centering
\resizebox{\hsize}{!}{\includegraphics{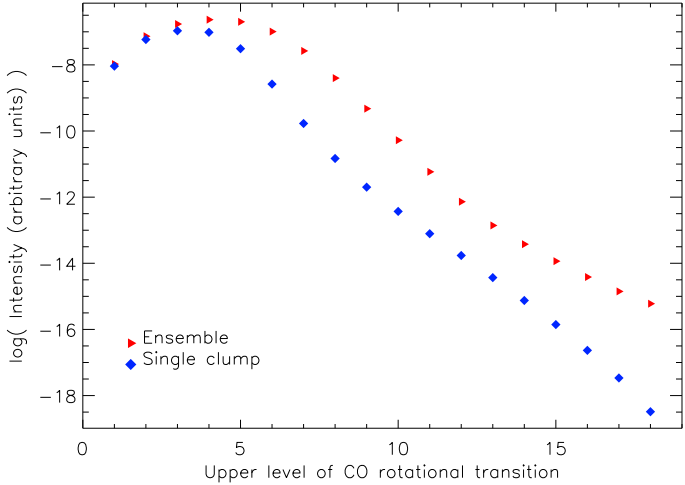}}
\caption{
 The \co~emission of a single clump of one solar mass 
 and a density of $n\ind{cl} = 10^{3.8}\ \pcc$ 
 compared to the emission 
 of an ensemble of clumps with the same total mass and 
 ensemble averaged clump density, embedded in an isotropic 
 FUV-flux of $10^{1.8}\ \fdraine$.
}
\label{f_ens-scl-comp}
\end{figure}

In Fig. \ref{f_ens-scl-comp} a direct comparison of the 
emission of an individual clump with the emission of a 
clump ensemble with 1 \msun{} is shown for the different 
\co~transitions. For $n\ind{ens} = 10^{3.8}\ \pcc$ 
and $\chi = 10^{1.8}\ \fdraine$ 
we find a significantly higher intensity in the mid-$J$ 
\co~transitions for the clump ensemble, which amounts 
to 2.5 orders of magnitude for the \co~8--7 line.
This is due to the fact, that the mid-$J$ \co~emission 
shows a steep decrease with clump mass (see Fig. 
\ref{f_mass-specific_emissivities}).

\begin{figure}
\centering
\resizebox{\hsize}{!}{\includegraphics{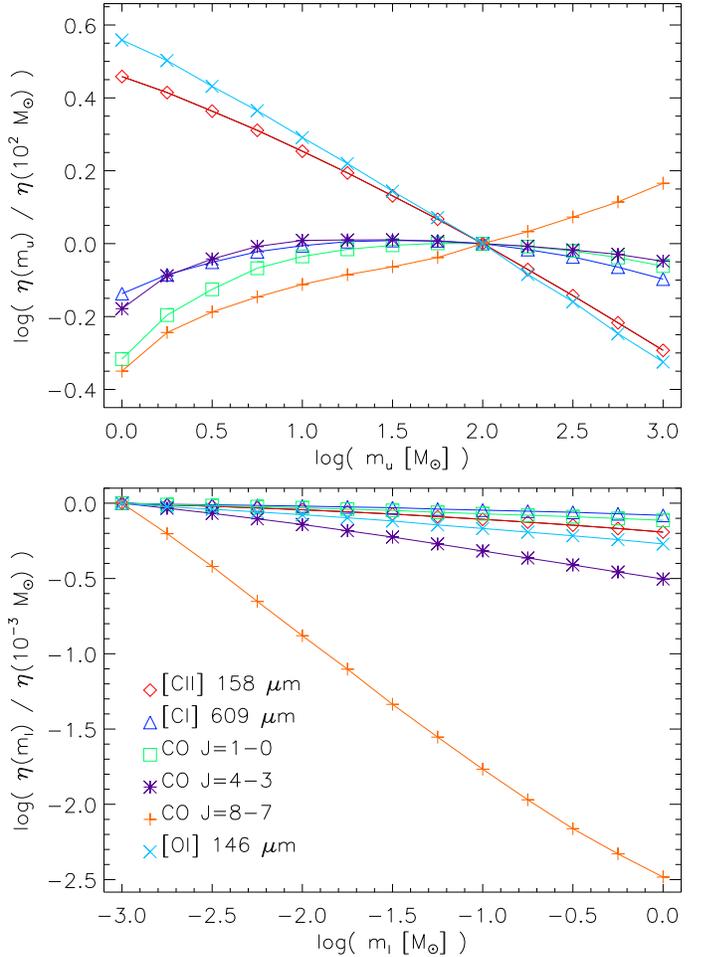}}
\caption{
 Dependencies of the clump ensemble emissivity $\eta$ on the 
 upper and lower clump mass, $m\ind{u}$ respectively $m\ind{l}$,
 for different tracers, $n\ind{ens} = 10^{3.8}\ \pcc$,
 and $\chi = 10^{1.8}\ \fdraine$.
 The values are normalized to the upper $m\ind{u}=10^2\ \msun$ respectively 
 lower clump mass limit $m\ind{l}=10^{-3}\ \msun$, as used in the model.
}
\label{f_clump_mass_limits}
\end{figure}

Fig. \ref{f_clump_mass_limits} illustrates the dependence of
the normalized emissivity with variation of the lower and upper
clump mass limits of the ensemble for different tracers.
Again, the mid-$J$ \co~emission shows a strong dependence 
on the lower clump mass limit, corresponding to the strong 
intensity enhancement with decreasing clump mass in these 
tracers. 
The \co~8--7 emissivity is nearly proportional to the lower 
clump mass limit.
The \cii{} and \oi{} emissivities show variations of 
one order of magnitude over an upper clump mass limit range 
of three orders of magnitude.
This is due to the increasing surface
to volume ratio of the clump ensemble with decreasing
upper clump mass limit (cf. Section \ref{s_clump_dist}) 
and the fact that the \cii{} and \oi{}~emissions arise 
mainly in shells in the outer parts of the clumps.

\section{Model results and comparison with observations}
\label{s_results}

\begin{figure}
\centering
\resizebox{\hsize}{!}{\includegraphics{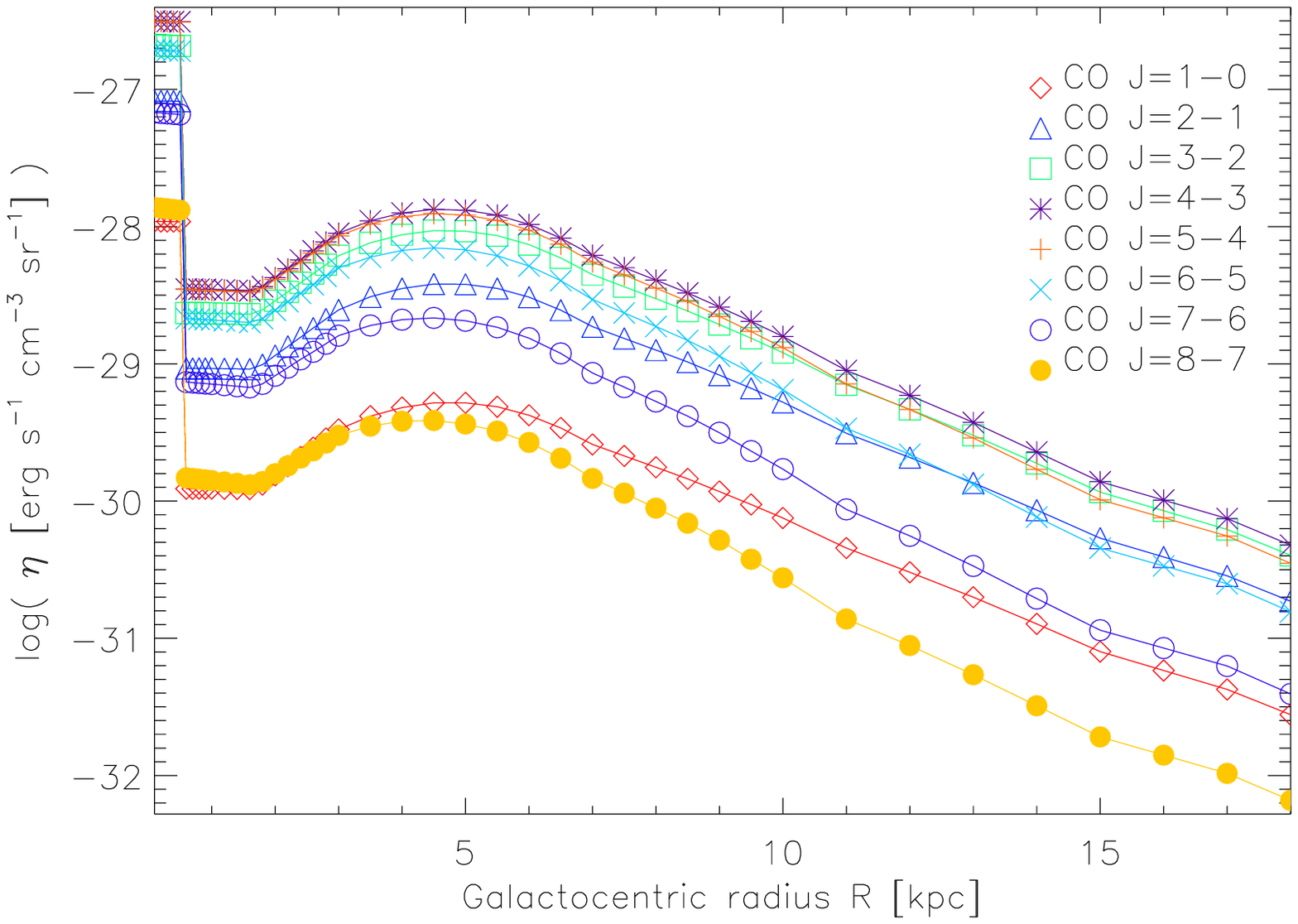}}
\caption{
 Calculated \co~emissivity distribution vs. Galactocentric radius 
 for $n\ind{ens} = 10^{3.8}\ \pcc$ and $\chi = 10^{1.8}\ \fdraine$
 at the solar circle.
}
\label{f_emissivity}
\end{figure}

The resulting emissivity distribution inherits the
intensity dependence of the clumps on 
the Galactic 
mass and density distributions as shown in Fig. 
\ref{f_emissivity}.

\begin{figure}
\centering
\resizebox{\hsize}{!}{\includegraphics{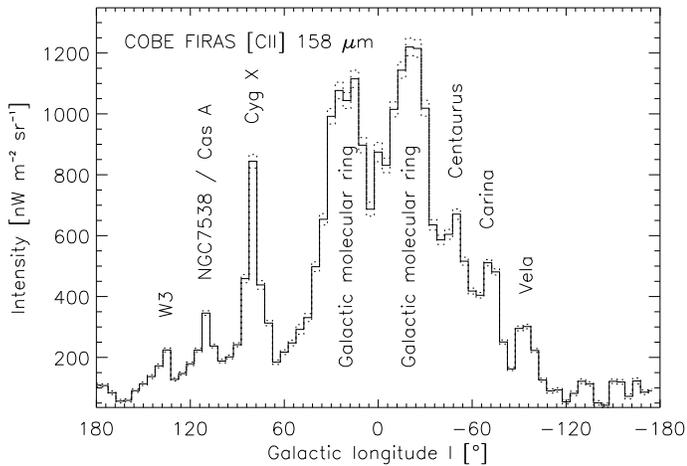}}
\caption{
 The \cii~158 \mum~intensity distribution with 
 Galactic longitude as published by \cite{Fixsen1999}. 
 The annotated source names are identified 
 using the \co~1--0 map published by \cite{Dame2001}.
}
\label{f_ann_cii_int}
\end{figure}


The beam averaged intensity is
determined according to the derivation in Section
\ref{s_emint}.
For the comparison of our model result 
with the COBE FIRAS observations 
we calculate the intensities in a beam with 
an extent of 5$^\circ$ in Galactic longitude
and 1$^\circ$ in Galactic latitude 
\citep{Fixsen1999}. 
All main PDR cooling lines in the COBE FIRAS
spectral range including \cii\ 158 \mum, 
\ci\ 609 and 370 \mum, \co\ $J=1$--0 to $J=8$--7, 
and \oi\ 146 \mum\ are used for the comparison.

To overview the emission features along the Galactic plane
we labeled the main features visible in the \cii{} intensity cut in
Fig. \ref{f_ann_cii_int}.
One can easily identify the Galactic large scale structure, 
dominated by the spiral arms and the Galactic molecular ring.


\subsection{Quantitative comparison}
\label{s_quantitative_comparison}



\begin{figure}
\centering
\resizebox{\hsize}{!}{\includegraphics{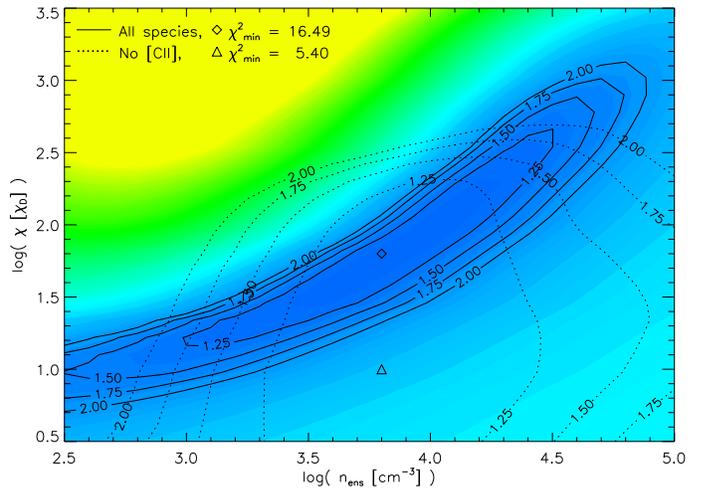}}
\caption{
 The reduced $\chi^2$ plotted as contours in the 
 ensemble averaged clump density-FUV-flux-plane. 
 The contours are given as factors of the minimum $\chi^2$ value.
 The solid line as well as the color code represents the $\chi^2$ 
 distribution including all species, while the dotted lines are the 
 $\chi^2$ contours neglecting the \cii\ intensity.
 The $\chi^2\ind{min}$ minima are given by the diamond and triangle 
 symbols, respectively.
}
\label{f_chi_contour}
\end{figure}

To quantify the quality of the model result 
we derive the reduced $\chi^2$ of the
intensity distributions.
The $\Delta I_i$ are the uncertainties of the observed
intensities due to the calibration of the COBE FIRAS instrument
and the data reduction.
The number of degrees of freedom is given by 
$ N\ind{dof} = N\ind{l} \, N\ind{spec} - 2 $,
with $N\ind{l}$ the number of bins in Galactic
longitude and $N\ind{spec}$ the number of species
used for the fit. The subtraction of 2 is due to
the 2 fitted parameters $n\ind{ens}$ and $\chi$.
Hence, the reduced $\chi^2$ is given by
\begin{equation}
\chi^2 = \frac{1}{N\ind{dof}} 
 \sum_{i=1}^{N\ind{l} \, N\ind{spec}} \left( \frac{ I_{\mathrm{obs},i} - I_{\mathrm{mod},i} }
 { \Delta I_i } \right)^2.
\end{equation}
The resulting $\chi^2$-distribution of the intensity 
distributions is plotted as 
contours onto the density-FUV-plane in Fig. 
\ref{f_chi_contour}.
The minimum $\chi^2$ has a value of 16.5
at an ensemble averaged clump density at the 
solar circle of
$n\ind{ens} = 10^{3.8}\ \pcc$ and 
an FUV-flux of $\chi = 10^{1.8}\ \fdraine$.
The $\chi^2$ distribution obviously confines
the density to FUV-flux ratio $n\ind{ens} / \chi$
very well, as this ratio determines the
molecular gas fraction within a PDR
\citep{Hollenbach1999}.
In contrast, the product $n\ind{ens} \chi$ is
hardly constrained.

%
%

\subsection{Discussion}

\begin{figure}
\centering
\resizebox{\hsize}{!}{\includegraphics{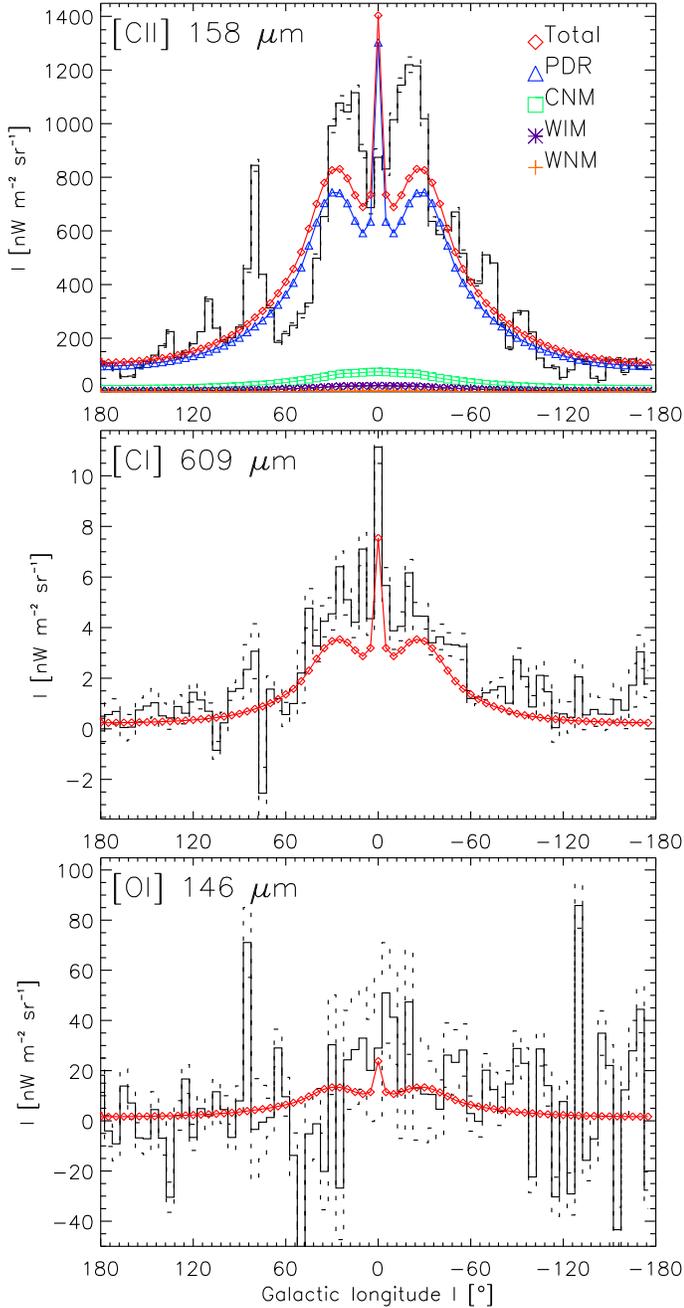}}
\caption{
 The model results of the \cii~158 \mum, the \ci{} 609 \mum{}, and 
 the \oi~146 \mum{} intensity distributions with 
 Galactic longitude overlaid on the observed distributions. 
}
\label{f_atomic_int}
\end{figure}

\begin{figure*}
\centering
\includegraphics[width=16cm]{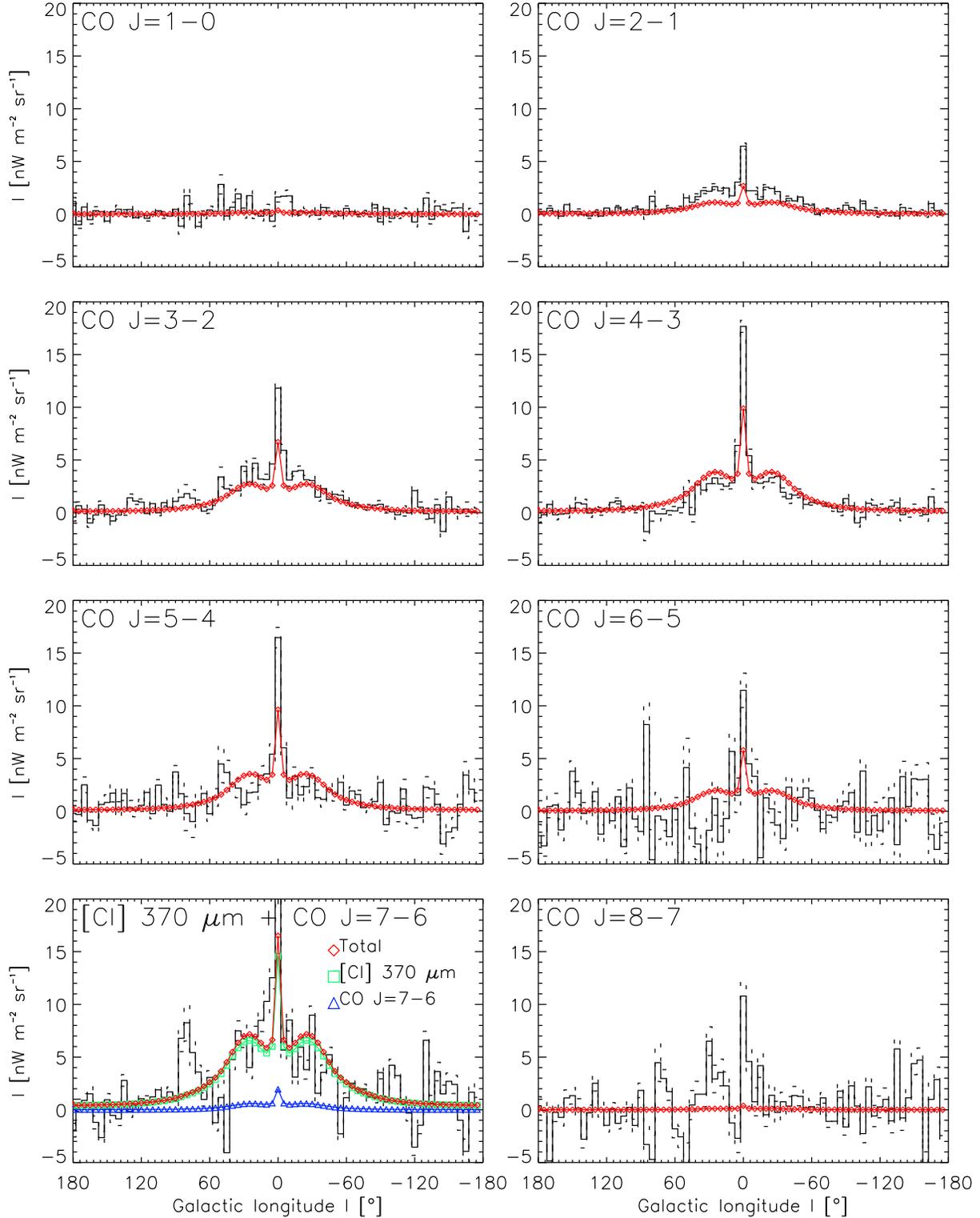}
\caption{
 The modeled intensities of the rotational transitions 
 $J=1$--0 (upper left) to $J=8$--7 (lower right) of \co{} overlaid
 the observational distributions along the Galactic plane.
 Note that the combined \co~$J=7$--6 and \ci~370 \mum{} line
 intensities are shown in the lower left as they are not
 spectrally resolved by the COBE FIRAS instrument.
}
\label{f_CO_int}
\end{figure*}

The results of the modeled line intensities are overlaid onto 
the observed intensities in Figs. 
\ref{f_atomic_int} and \ref{f_CO_int}.
We note that the relatively high value of
the minimum $\chi^2$ is due to the pronounced
Galactic large scale structure in the observed
\cii\ emission (which is not contained in the 
smooth model distribution)
and the high signal to noise ratio of these data. 
In fact, it is surprising that one can find average
values for these parameters, which allow us to reproduce
the bulk of the FIR line emission of the Milky Way.
The partly significant local deviations of the model
results from the observations are due to the strongly
simplifying assumptions. 
Except \co~$J=1$--0, 6--5, and 8--7, which have large
observational uncertainties, all lines are
reproduced in total Galactic flux within factors below
2.3 (\co\ $J=2$--1).
The mismatch of the \co~2--1 indicates an additional cold
molecular cloud component in the Galactic gas distribution.

The resulting \thirteenco~1--0 / \co~1--0 line 
intensity ratio with a value of about 4 is in accordance with
observational results \citep{Gierens1992, Falgarone1998}.

\subsubsection{Contributions to the Galactic \cii{} 158 \mum{} emission}
\label{s_cii-discussion}

We now turn to the comparison of the COBE observed \cii{} intensities with the
model results (see Fig. \ref{f_atomic_int}). 
Additional flux is measured mainly from the Galactic ring, 
the Cyg X region and other star-forming regions in the spiral arms. 
This is in contrast to the picture of \cii-emission stemming to a 
large part from the CNM \citep{Hollenbach1999}.
Based on direct UV absorption measurements of the
excited state of \cplus{}, \citet{Pottasch1979} and \citet{Gry1992} 
derive an average emissivity of the CNM, in 
particular diffuse \hi-clouds, of 
$5 \times 10^{-26}$ erg~\ps~H-atom$^{-1}$. Using this number the
contribution from PDRs to the Galactic \cii-emission should be very small.
This has been questioned by \citet{Petuchowski1993}
who accounted for about half of the \cii-emission from PDRs,
the other half from the WIM and only a very small fraction
from the CNM. In contrast, \citet{Heiles1994} found that the CNM
ranks second after the WIM while PDR contributions should be small.
\citet{Carral1994} state that up to 30\% of the
interstellar \cii~emission may stem from \hii-regions.
\citet{Kramer2005} used ISO/LWS observations of the \nii~122 \mum{}
line in the spiral arms of M83 and M51 to estimate that between 15\%
and 30\% of the observed \cii~emission originates from \hii-regions.
Computations by \citet{Abel2005} indicate that the main
\cii-emission arises from PDRs and a fraction of 10\% up to
60 \% from \hii-regions.
Calculations by \citet{Kaufman2006} find similar
\cii-emission fractions, deviating by less than a factor of 2.

We can test this result from the viewpoint of the excitation of 
\cplus{} which is a simple problem because the fine structure
line arises basically from a two-level system. Using the 
excitation rates for collisions with H, H$_2$, He, and electrons
from \citet{Flower1977}, \citet{Launay1977}, and \citet{Wilson2002} 
we can compute
an upper limit to the emissivity of C$^+$ as a function of density
under the temperature and ionization conditions of the different 
phases of the ISM via
\begin{equation}
\eta_{\rm esc} = \frac{1}{4\pi} \frac{ n_{\rm C^+} A_{\rm ul} 
 E_{\rm ul} \beta }{ 1 + \frac{g_{\rm l}}{g_{\rm u}} 
 \exp \left( \frac{E_{\rm ul}}{\rm{k}_{\rm B} T} \right)
 (1 + \frac{n_{\rm cr} \beta}{n}) }
\label{e_esc_prob}
\end{equation}
with $n$ the total number density of the collision partners 
(mainly hydrogen atoms or electrons), 
$A\ind{ul}$ the Einstein coefficient for spontaneous emission, 
$g\ind{l,u}$ the statistical weights for the lower and upper state, 
$n\ind{\cplus}$ the number density of \cplus, 
$E\ind{ul}$ the energy spacing between the upper and lower state, 
$n\ind{cr}$ the critical density of the 
\cii~\element[][2][][3/2]{P}-\element[][2][][1/2]{P}
transition with respect to neutral particles or 
$n_{\rm cr,e}$ for electrons, 
k$_{\rm B}$ the Boltzmann constant, 
and $T$ the temperature.
The escape probability $\beta$ describes the probability
of a photon of the considered transition to escape the
cloud from its position within, assuming 
constant excitation conditions allover the cloud.
To obtain an upper limit of the emission of the atomic 
gas components, we assume optically thin emission, i.e. 
$\beta = 1$.
The values for the different parameters are summarized 
in Table \ref{t_escprob}.
\begin{table}
\caption{The parameter values used for the escape
 probability calculation of the
 \cii~\element[][2][][3/2]{P}-\element[][2][][1/2]{P}
 emissivity from the CNM and WNM. Powers of ten are
 given in parentheses.
} 
\label{t_escprob}
\centering
\begin{tabular}{lr}
\hline \hline 
Quantity & Value \\
\hline 
$\beta$ & 1 \\
$g_{\rm u}$ & 4/3 \\
$g_{\rm l}$ & 2/3 \\
$A_{\rm ul}$ [\ps] & $2.4\,(-6)$ \\
$n_{\rm cr}$ [\pcc] & $4\,(3)$ \\
$n_{\rm cr,e}$ [\pcc] & $4\,(1)$ \\
\hline
\end{tabular}
\end{table}

\begin{figure}
\centering
\resizebox{\hsize}{!}{\includegraphics[angle=90]{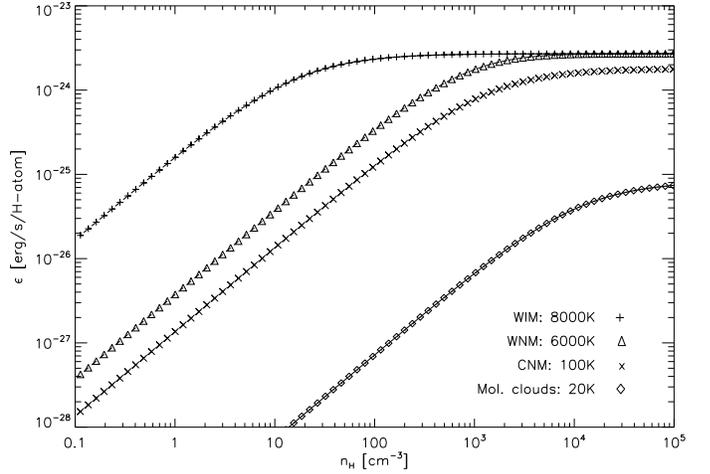}}
\caption{Optically thin emissivity of \cii{} under the excitation conditions
of the different ISM components as a function of the gas density. 
}
\label{c+_contributions}
\end{figure}

Fig. \ref{c+_contributions} shows the \cii~emission 
from the different phases.
All carbon is considered to be singly ionized, using an elemental
abundance of $1.4 \times 10^{-4}$ (see Table \ref{t_kosma_par}). 
Upper limits are also guaranteed by assuming an
ionization degree, i.e. an electron density relative to the total 
density of protons, of $10^{-3}$ in the CNM, $10^{-2}$ in the WNM,
and 1 in the WIM (see Table \ref{t_galpar} for the local Galactic 
parameter values of the different phases).
We find that the ``standard value'' of 
$5 \times 10^{-26}$~erg~\ps~H-atom$^{-1}$ 
from \citet{Hollenbach1999} is in 
agreement with the density range of the 
CNM given in \citet{Wolfire2003}.
At low densities an efficient excitation 
of \cplus{} is only possible 
if a large fraction of the collision 
partners is charged, i.e.
in the ionized medium.

\begin{figure}
\centering
\resizebox{\hsize}{!}{\includegraphics{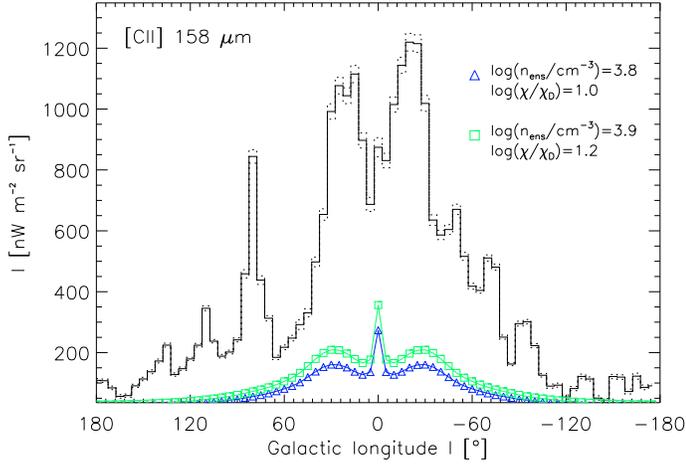}}
\caption{
 PDR \cii\ intensity corresponding to the best fit parameter values 
 ignoring the \cii\ emission and under the assumption that the contribution 
 of the PDR-correlated \hii\ regions to the total \cii\ emission is 60\%.
}
\label{f_cii_parvar}
\end{figure}

The results of \citet{Petuchowski1993} and
\citet{Abel2006} who found, that the 
\cii~and \nii~emission are strongly correlated, 
indicate that main contributions (60\% maximum)
are expected from low density \hii-regions.
Using this result, we rerun the $\chi^2$ fit
of the lines but increase 
the computed \cii-intensity by a global factor 2.5 
before fitting the model results to the observations.
This fit provides a lower limit of the FUV-flux of 
$\chi = 10^{1.2}\ \fdraine$,
which is slightly below the estimated 
range from Section \ref{s_galpar_dist} with a lower
limit of $\chi = 10^{1.3}\ \fdraine$.
The best fitting ensemble averaged clump density is 
derived to $n\ind{ens} = 10^{3.9}\ \pcc$ in this case.
Additionally we check the fit results completely 
ignoring the \cii\ intensity distribution. 
In this case we obtain
parameter values of $n\ind{ens} = 10^{3.8}\ \pcc$ and 
$\chi = 10^{1.0}\ \fdraine$. 
The $\chi^2$-distribution
with a minimum of 5.40 for this case is also 
shown in Fig. \ref{f_chi_contour}.
The resulting low FUV-flux is due to the low
signal to noise ratio in the \oi\ and upper
\co\ transitions, so that this result is biased
by the low-$J$ \co\ and \ci\ emission and lies
below our estimate for the average 
FUV-flux seen by the Galactic molecular gas.
The confinement of the $\chi^2$ distribution
ignoring the \cii\ intensity distribution is
weak in either direction. In particular at 
$n\ind{ens} = 10^{3.8}\ \pcc$ and 
$\chi = 10^{1.8}\ \fdraine$
we find a $\chi^2$ value of 5.75 -- only about 7\% 
more than the minimum value, indicating that the
\cii\ emission fixes the resulting FUV-flux value
of the model.
The computed PDR \cii\ intensity distributions for 
these cases are shown in Fig. \ref{f_cii_parvar}.

There are smaller, comparable contributions from 
the diffuse WIM, which has an average density well 
below 1 \pcc{} \citep{McKee_Ostriker1977, Reynolds1991, 
Cox2005, Hill2007}, 
and from the CNM , and a vanishing contribution of the WNM.
However, the much larger scale height of the diffuse WIM 
leads to a reduced contribution in comparison to the CNM, 
in agreement with our result for the intensity distribution 
along the Galactic plane shown in Fig. \ref{f_atomic_int}. 

We conclude that our model can reproduce the Galactic 
\cii\ emission with the main contribution from clumpy
PDRs, but we cannot exclude a major contribution 
from low-density \hii\ regions.

\subsubsection{The FIR continuum emission}

\begin{figure}
\centering
\resizebox{\hsize}{!}{\includegraphics{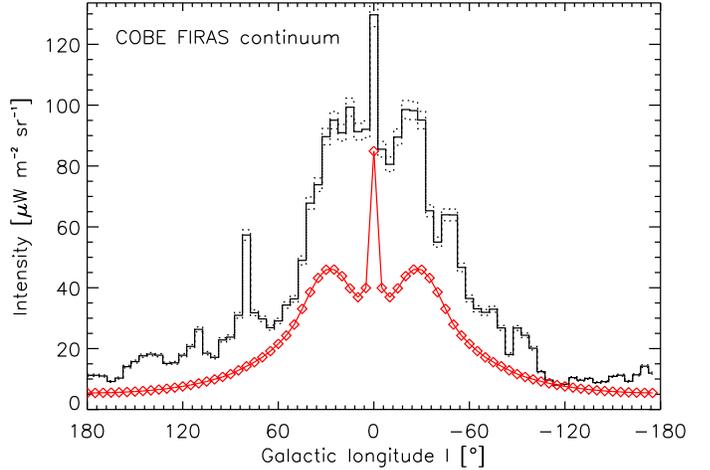}}
\caption{
 The modeled FIR intensity distribution along the Galactic
 plane in the COBE FIRAS wavelength band overlaid on the 
 observed distribution as published by \cite{Fixsen1999}.
}
\label{f_int_FIR}
\end{figure}

Only 51\% of the Galactic FIR emission in the COBE FIRAS 
spectral range is reproduced by a separate dust radiative 
transfer calculation for the \kosma~clumps (R. Szczerba, 
priv.\ comm.) as shown in Fig. \ref{f_int_FIR}.
This is expected, as the dust is also heated by less energetic
radiation than the FUV from the older stellar population.
\citet{Mochizuki2000} argue that about half of the total stellar
flux lies in the FUV spectral range in agreement with our result.


\section{Summary and Conclusions}

The bulk of the FIR and submillimeter line emission in 
\cii~158 \mum, \ci~609 \mum{} and 370\mum, \oi~146 \mum, and 
\co~1--0 to 8--7 of the Milky Way as 
observed by COBE FIRAS can be reproduced by a clumpy
PDR-model, which takes the fractal structure of the dense ISM 
as a clump ensemble with a power law clump mass and
size distribution into account.
This is a remarkable result, as the model
contains essentially no free parameters; 
all parameters being constrained by independent knowledge 
about the structure, kinematics of, and star formation rate 
in the disk of the Milky Way.

The result suggests that the bulk of the 
Galactic FIR line emission stems from fractally structured
PDRs, illuminated by the FUV radiation field emitted 
by the Galactic population of massive stars in
adjacent star forming regions. 

The UV-penetrated, clump-cloud PDR scenario applied here 
to the Milky Way line emission as observed by COBE is a 
versatile approach, which can be applied in the future 
to model the emission from individual Galactic star 
forming regions as well as the line maps of spatially 
resolved nearby galaxies.

Future studies and observations should address the 
validity of particular assumptions, like
the limits of the clump mass distribution.
Observations with high spatial and/or spectral resolution, 
that will be possible with upcoming missions as SOFIA, 
Herschel, and ALMA, will give further insights in the
nature of the ISM.


\begin{acknowledgements}
 We are grateful to D. J. Fixsen for providing us the results of 
 \cite{Fixsen1999} in electronic form.
 We thank the anonymous referee for her/his helpful comments.
 This work is supported by the Deutsche Forschungsgemeinschaft
 (DFG) via Grant SFB 494.
\end{acknowledgements}

\bibliography{9270}
\bibliographystyle{aa}


\end{document}